\documentclass[12pt]{article}
\usepackage{epsfig}
\usepackage{graphicx}
\usepackage{amsmath}
\usepackage{amssymb}
\def\Title#1{\begin{center} {\Large {\bf #1} } \end{center}}

\begin{document}

\Title{Proposal for Single-Bunch Collimator Wakefield Measurements at SLAC ESTB}

\bigskip\bigskip


\begin{raggedright}  

{\it A.~Latina, G.~Rumolo, D.~Schulte, R.~Tomas}\\
CERN, Geneva, Switzerland\\
{\it R.~M.~Jones}\\
Cockcroft Institute, Daresbury, UK\\
{\it J.~Smith}\\
Tech-X UK Ltd, UK\\
{\it C.~I.~Clarke, C.~Hast, M.~Pivi}\\
SLAC, Stanford, USA\\
{\it A.~Faus-Golfe, N.~Fuster-Martinez, J.~Resta-Lopez}\footnote{resta@ific.uv.es}\\
IFIC (CSIC-UV), Valencia, Spain
\bigskip\bigskip
\end{raggedright}

\begin{abstract}
Collimator wakefields in the Beam Delivery System (BDS) of future linear colliders, such as the International Linear Collider (ILC) \cite{ILC} and the Compact Linear Collider (CLIC) \cite{CLIC}, can be an important source of emittance growth and beam jitter amplification, consequently degrading the luminosity. Therefore, a better understanding of collimator wakefield effects is essential to optimise the collimation systems of future linear colliders in order to minimise wakefield effects. In the past, measurements of single-bunch collimator wakefields have been carried out at SLAC \cite{experi1,  experi2, experi3, experi4} with the aim of benchmarking theory, numerical calculations and experiments. Those studies revealed some discrepancies between the measurements and the theoretical models. New experimental tests using available beam test facilities, such as the End Station A Test Beam (ESTB) at SLAC \cite{ESTBSLAC}, would help to improve our understanding on collimator wakefields. ESTB will provide the perfect test bed to investigate collimator wakefields for different bunch length conditions, relevant for both ILC (300~$\mu$m nominal bunch length) and CLIC (44~$\mu$m nominal bunch length) studies. Here we propose to perform new experimental tests of collimator wakefield effects on electron/positron beams at SLAC ESTB. 
\end{abstract}

\section{INTRODUCTION}
Collimator wakefields in the Beam Delivery System (BDS) of the future linear colliders are expected to be an important source of emittance growth and beam jitter amplification, consequently degrading the luminosity. In order to alleviate the collimator wakefield effects, a lot of efforts are being dedicated to the optimisation of the collimator systems of the linear colliders, regarding the geometric design and the material selection. 

\begin{figure*}[htb]
\centering
\includegraphics[width=78mm]{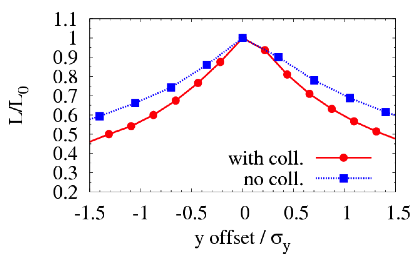}
\includegraphics[width=70mm]{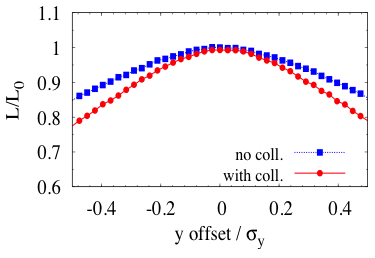}
\caption{Luminosity loss with and without collimator wakefield effects for the ILC (left) and
CLIC (right), assuming nominal beam parameters. } \label{lumiplot}
\end{figure*}

Figure~\ref{lumiplot} illustrates the luminosity degradation due to collimation wakefield effects for both the ILC and CLIC, based on simulations using the codes PLACET \cite{PLACET} and GUINEAPIG \cite{GUINEAPIG}, and assuming nominal parameters. For instance, considering an incoming beam with $0.5~\sigma_y$ offset with respect to the beam axis, taking into account the collimator wakefield effects one obtains about $20\%$ luminosity loss in comparison with $15\%$ luminosity loss without wakefield effects. 

Collimator wakefield measurements are relevant to understand and evaluate the real effects on the beam. This helps to improve the theoretical and numerical models for a more realistic estimate of the machine performance. 

In the past, sets of measurements have been made for longitudinally tapered collimators at SLAC, see e.g.~\cite{experi1, experi2, experi3, experi4}. For the geometric wakefields, these measurements showed an agreement at the level of $20\%$ with the simulation results and good qualitative agreement with the theory, although in many cases there was a quantitative discrepancy as large as a factor 2 between theory and measurement. Measurements of the resistive wakefields    \cite{experi4} showed notable discrepancies with theory. No precise measurement of the bunch length was performed during those collimator wakefield tests, and this lack of bunch length information could be the main source of discrepancy between measurements and theoretical models. Further investigation is needed at this respect. 

End Station A Test Beam (ESTB)  \cite{ESTBSLAC}, former End Station A (ESA), will provide the necessary test beams and an excellent experimental environment for performing new sets of measurements of collimator wakefields. This will be very helpful for a better understanding of the collimator wakefield effects.

Table~\ref{tableparameters} shows some relevant parameters for ESTB, in comparison with the nominal parameters for the BDS of ILC (at 0.5~TeV centre-of-mass energy) and CLIC (at 3~TeV centre-of-mass energy). 

\begin{table}[htb]
\begin{center}
\caption{ESTB primary electron beam parameters, compared with the nominal beam parameters for ILC and CLIC.}
\begin{tabular}{|l|c|c|c|}
\hline \textbf{Parameter} & \textbf{ESTB} & \textbf{ILC BDS} & \textbf{CLIC BDS} \\
\hline Beam energy [GeV] & 15 & 250 & 1500 \\
\hline Repetition rate [Hz] & 1--5 & 5 & 50 \\
\hline Energy spread [$\%$] & 0.02 & 0.1 & 0.3 \\
 \hline Bunch charge [nC] & 0.35 & 3.2 & 0.6 \\
 \hline Bunch length (rms) [$\mu$m] & 100 & 300 & 44 \\
 \hline Normalised emittance & & & \\
 ($\gamma\epsilon_x$, $\gamma\epsilon_y$) [$\mu$m-rad] & (4, 1) & (10, 0.04)  & (0.66, 0.02) \\
\hline
\end{tabular}
\label{tableparameters}
\end{center}
\end{table}

\section{BRIEF THEORETICAL BACKGROUND}

\subsection{Geometric Wakefields}
\begin{figure*}[htb]
\centering
\includegraphics[width=85mm]{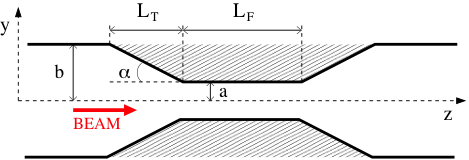}
\includegraphics[width=65mm]{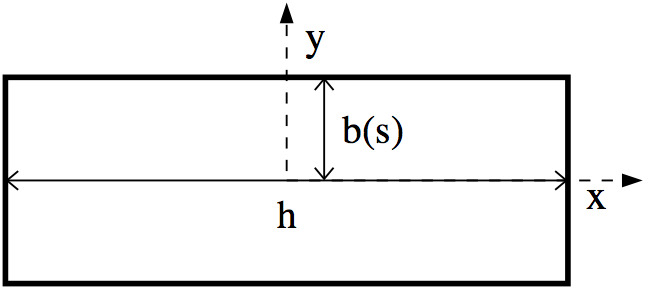}
\caption{Schematic longitudinal view (left) and cross-sectional view (right) of a rectangular tapered collimator.} \label{collscheme}
\end{figure*}

There are analytical approximate expressions that allow a quick prediction of the expected transverse linear wake kick factors, knowing the collimator characteristics and the beam parameters.  Following the StupakovÕs prescriptions \cite{Stupakov}, the geometrical wake kick factor ($\kappa_g$) can be calculated using the following Ònear-centreÓ approximation for rectangular collimators in different regimes, depending on the ratio between the bunch length $\sigma_z$ and the geometrical features of the collimator: 

\begin{equation}\label{eq1}
\kappa_g=\frac{Z_0 c}{4\pi} 
\begin{cases}
\sqrt{\pi}\alpha h/(4 \sigma_z)\left( 1/a^2 - 1/b^2 \right) & \text{for } \sqrt{\alpha a / \sigma_z }< 6.2 a /h; \qquad \quad \,\,\, \text{inductive} \\
8/3 \sqrt{\alpha /(\sigma_z a^3)} & \text{for } 0.37 > \sqrt{\alpha a / \sigma_z} > 6.2 a /h; \,\, \text{intermediate} \\
1/a^2 & \text{for } \sqrt{\alpha a / \sigma_z} > 0.37; \qquad \qquad \,\,\, \text{diffractive}
\end{cases}
\end{equation}

\noindent where $b$ and $a$ denote the maximum and minimum half gap of the collimator, respectively; $\alpha$ represents the collimator taper angle, and $h$ the full width of the gap in the non-collimating direction (see Fig.~\ref{collscheme} for definition of collimator parameters). $Z_0 = 376.7~\Omega$ is the impedance of free space and $c$ the speed of light.

\begin{figure*}[htb]
\centering
\includegraphics[width=120mm]{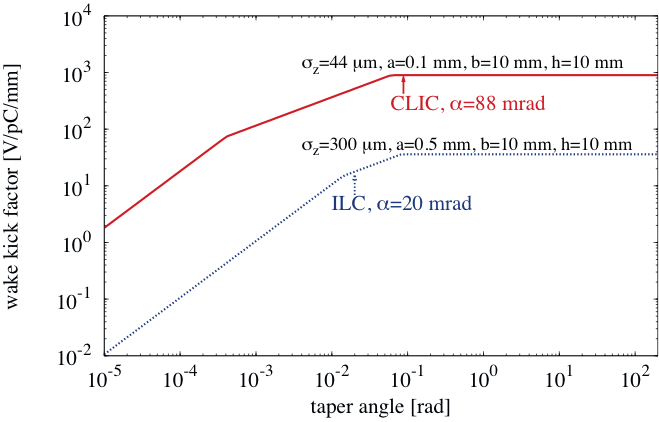}
\caption{Geometric wake kick factor versus taper angle considering ILC and CLIC collimator parameters.} \label{plotregimes}
\end{figure*}

For the sake of comparison, Fig.~\ref{plotregimes} shows the geometric kick factor $\kappa_g$ as a function of the taper angle $\alpha$, considering the parameters for the betatron collimators of ILC and CLIC.

\subsection{Resistive Wakefields}
The transverse kick factor corresponding to the resistive-wall wakefield ($\kappa_r$) of rectangular tapered collimators can be approximated by the following expressions for very small beam offsets:

\begin{itemize}
\item Long range regime \cite{Piwinski}, $0.63\left(2a^2/(Z_0\sigma)\right)^{1/3} \ll \sigma_z \ll 2 a^2 Z_0 \sigma$: 

\begin{equation}\label{eq2}
\kappa_r \simeq \frac{Z_0 c}{4\pi} \frac{\pi}{8a^2} \Gamma \left(1/4\right)\sqrt{\frac{2}{\sigma_z \sigma Z_0}}\left[\frac{L_F}{a} + \frac{1}{\alpha}\right]\,\,,
\end{equation}

\noindent  where $\sigma$ is the electrical conductivity of the collimator material, $L_F$ the length of the flat part of the collimator, and $\Gamma (1/4) = 3.6256$.

\item Short range regime \cite{Bane}, $\sigma_z <0.63 \left(2a^2 /(Z_0 \sigma)\right)^{1/3}$:

\begin{equation}\label{eq3}
\kappa_r= \frac{Z_0 c}{4\pi} \langle W^1_{\perp} \rangle \,\,,
\end{equation}

\noindent where $W^1_{\perp}$ is the transverse dipole wake potential:

\begin{equation}\label{eq4}
W^1_{\perp}(s)=\int_0^{\infty} w^1_{\perp} \rho_{z} (s-s') \, \mathrm{d}s' \,\,,
\end{equation}

with $\rho_z$ the longitudinal bunch distribution and $w^1_{\perp}$  the dipole wake function $w^1_{\perp}=| {\bf w}_{\perp} (r,s)|/r$  (with $r$ the transverse offset of the driving charge), which can be calculated using the Panofsky-Wenzel theorem \cite{PanofskyWenzel},

\begin{equation}\label{eq5}
\frac{\partial \mathbf{w}_{\perp}}{\partial s}=\vec{\nabla}_{\perp} w_z\,\,,
\end{equation}

\noindent if the longitudinal wake function $w_z$ is known. K. Bane and M. Sands \cite{Bane} calculated $w_z$ for the short-range regime: 

\begin{equation}\label{eq6}
w_z(s)=\frac{16}{b(s)^2}\left[ \frac{e^{-s/s_0}}{3}\cos\frac{\sqrt{3}s}{s_0}-\frac{\sqrt{2}}{\pi}\int_0^{\infty} \frac{x^2 e^{-x^2 s/s_0}}{x^6 + 8} \mathrm{d}x\right]\,\,,
\end{equation}

\noindent where $s$ is the distance the test particle is behind the driving charge, and $s_0 (s)= (2b(s)^2/(Z_0\sigma))^{1/3}$, with $b(s)$ the changing radius from the maximum aperture to the minimum aperture in the collimation plane.

\end{itemize}

\section{EXPERIMENTAL CONCEPT}
\begin{figure*}[htb]
\centering
\includegraphics[width=150mm]{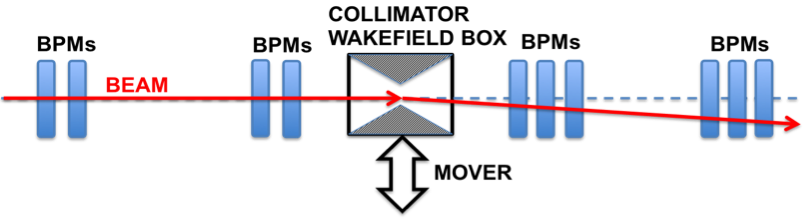}
\caption{Simple schematic showing the configuration of BPMs for wakefield kick measurement.} \label{experischeme}
\end{figure*}

The experimental concept, schematically illustrated in Fig~\ref{experischeme}, is relatively simple. The collimator jaws to be tested are placed in a drift space of the beamline. The beam trajectory is reconstructed using BPMs upstream and downstream of the collimator.  

For each set of collimator jaws the beam is centred through the collimator using upstream magnet corrector or position feedbacks. Then a reference orbit is recorded without moving the collimator. 

The collimator is located in a metallic box equipped with vertical movers. This allows to move the collimator vertically around the beam and to make a vertical position scan. The reference orbit is subtracted from the data taken during the vertical position scan, in order to suppress the DC offsets of the BPMs. The direct result of these measurements is the deflection angle of the beam as a function of the beam-collimator offset. A statistical error is determined for each deflection. The linear fit to these data gives the linear wake kick, which is the figure of merit in this study.

\section{EXPERIMENTAL EQUIPMENT}
\subsection{Collimator Prototypes}
\begin{figure*}[htb]
\centering
\includegraphics[width=110mm]{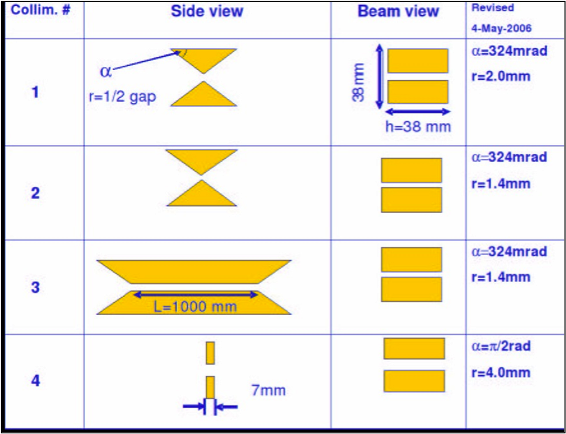}
\caption{Geometrical structure of one of the sets of collimators used for the wakefield measurements at SLAC ESA. These collimators are made of copper. Figure from [5], courtesy of J. L. Fernandez-Hernando. } \label{collprototypes}
\end{figure*}

16 different collimator configurations of geometry and material were tested in 4 test beam runs at ESA during 2006 and 2007 (T-480 experiment) \cite{experi3}. An example of the geometrical structure of a set of four of these collimators is shown in Fig.~\ref{collprototypes}. These collimator prototypes were manufactured under the auspices of the Science and Technology Facilities Council (STFC) from UK, and recently they have generously been lent to CERN for new studies. Following the collimator specifications for the CLIC collimation system (see the CLIC CDR \cite{CLIC}), new collimator prototypes could be manufactured.  

The goal is to study the wakefield dependence on different geometric collimator parameters, e.g. jaw apertures, lengths and tapering angle, as well as on the collimator material properties. This will allow us to investigate optimal materials and geometry configurations to minimise wakefields.  

\subsection{Collimator Wakefield Box}
\begin{figure*}[htb]
\centering
\includegraphics[width=120mm]{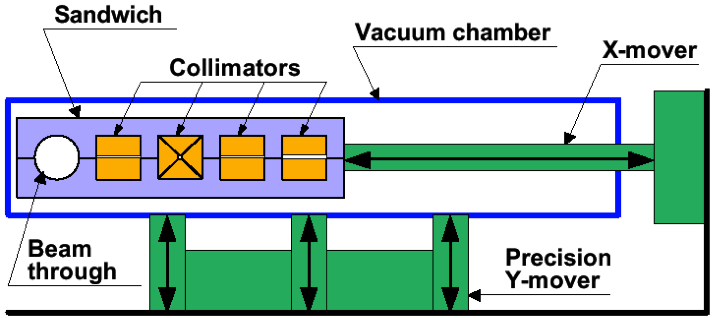}
\caption{Schematic of collimator wakefield beam test apparatus.} \label{collwakebox}
\end{figure*}

This apparatus consists of a rectangular stainless steel vacuum chamber containing five possible slots for the beam to move through. Four of the slots are occupied with collimators and the fifth is empty to allow normal operation. The box could be pushed horizontally by a mover in order to change the slot and therefore the collimator presented to the beam. This box is also controlled by three movers for vertical translations in 1~$\mu$m steps, and allows measuring the change in the kick imparted to the beam for different offsets of the beam respect to the centre of the collimators. This test apparatus was designed for collimator wakefield measurements at SLAC, and it was used in previous experiments at ESA (T-480) \cite{Tenenbaum}. A cutaway schematic of this device is shown in Fig.~\ref{collwakebox}. This device has been already installed in ESTB (former ESA), and ready to be used for new experiments like the one proposed here.

\subsection{Beam Position Monitors}
\begin{figure*}[htb]
\centering
\includegraphics[width=95mm]{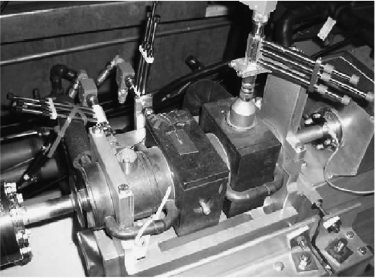}
\caption{Example of a rectangular cavity-type BPM used in the ESA beamline for the ILC energy spectrometer experiment \cite{ESAexperiment}.} \label{ESABPM}
\end{figure*}

In principle we plan to use available BPMs, recycled from old ESA experiments, such as the ILC energy spectrometer experiment \cite{ESAexperiment}. A typical rectangular cavity BPM used in ESA is shown in Fig.~\ref{ESABPM}. The ESA cavity BPMs demonstrated resolutions in the range $0.17~\mu$m -- $2.2~\mu$m, and the system was stable at the micron level over the course of 1~h.  

\begin{figure*}[htb]
\centering
\includegraphics[width=90mm]{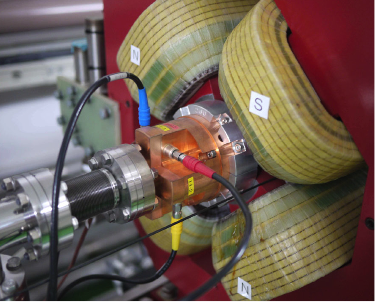}
\caption{Example of a C-band cavity BPM inside a quadrupole magnet of the ATF2 beamline. Image courtesy of Y. I. Kim.} \label{ATF2BPMplot}
\end{figure*}

For our experiment we can also consider the improvement of the system by the use of C-band BPMs like those used in the KEK Accelerator Test Facility 2 (ATF2) in Japan \cite{YIKim} (see Fig.~\ref{ATF2BPMplot}). The ATF2-like C-band cavities have demonstrated to be highly reliable, achieving resolutions in the range 30 nm -- 250 nm.  

Simulation studies to determine the required BPM resolution for the collimator wakefield experiment in ESTB are performed in Section 6.3. 

\subsection{Bunch Length Measurement}
An important ingredient of this experiment will be the precise measurement of the bunch length. The rms bunch length strongly influences wakefield kick factors. Therefore, reliable and systematic bunch length measurements will be essential for a more precise comparison of measured wakefields with theoretical calculations and simulations. 

For the reconstruction of the bunch time profile and bunch length measurement, different techniques could be used, e.g. RF deflector, Smith-Purcell radiation and electro-optic methods. 

\begin{figure*}[htb]
\centering
\includegraphics[width=120mm]{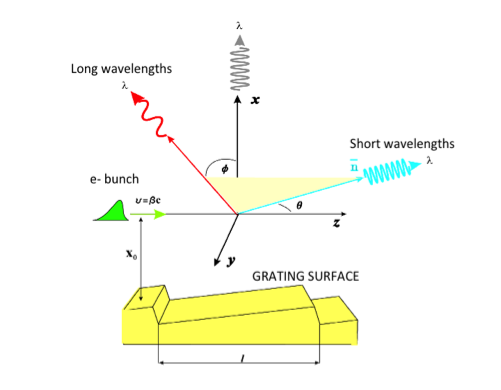}
\caption{Schematic of the Smith-Purcell radiative process.} \label{SPRscheme}
\end{figure*}

For this experiment we are planning to use a Smith-Purcell Radiation (SPR) bunch profile monitor, which is a very promising non-intercepting technique. It is based on the detection of the coherent Smith-Purcell radiation emitted when a charged particle beam passes near a metallic grating. The grating acts like a monochromator and disperses the radiation according to the observation angle. By measuring the intensity distribution it is possible to reconstruct the bunch time profile. Figure~\ref{SPRscheme} illustrates the SPR process, which is extensively explained in \cite{Doucas}. 

\begin{figure*}[htb]
\centering
\includegraphics[width=100mm]{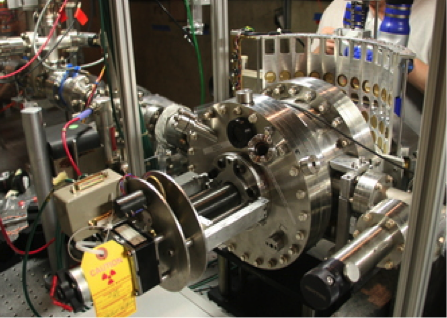}
\caption{Smith-Purcell Radiation monitor in the FACET beamline. Courtesy of G. Doucas.} \label{FACETSPR}
\end{figure*}

The SPR technique allows the bunch measurement in a wide range (from few fs to ps). In 2011 a SPR monitor was installed in FACET\footnote{FACET, Facility for Advanced Accelerator Experimental Test \cite{FACET}, is a new facility which uses two-thirds of the $3.2$~km long SLAC linac for supporting studies from many fields, but in particular those of plasma wakefield acceleration and dielectric wakefield acceleration. FACET provides high energy density electron and positron beams with peak currents of $\approx 20$~kA that are focused down to $\approx 10\times 10$ micron transverse spot size at an energy of $\approx 20$~GeV.} and is currently being tested (see Fig.~\ref{FACETSPR}). The aim is to measure bunch lengths in the $< 50~\mu$m scale ($< 0.2$~ps). This will be relevant to investigate collimator wakefields with CLIC-like bunches ($\sigma_z = 44~\mu$m). During the 2011--2012 experimental campaigns very promising results have been obtained with temporal bunch profiles (FWHM) of the order of 100~fs \cite{Bartolini}. 

Recently the collimator wakefield collaboration and the E-203 collaboration have agreed to join efforts to build up a second SPR monitor to be installed in ESTB in support of collimator wakefield studies. The goal is also to upgrade this device for single shot measurements (2nd generation device). The plan is as follows:

\begin{itemize}
\item Fix design by January 2013 based on improved 1st generation device.
\item Build up a second SPR monitor ready for use in September 2013 at ESTB.
\item Include all the accumulated learning from the FACET device that enables single shot capability.
\item Get as close as possible to a single shot device in the design stage.
\item And evolve in stages from averaging to single shot operation.
\end{itemize}

\section{ESTB OPTICS}
\begin{figure*}[htb]
\centering
\includegraphics[width=120mm]{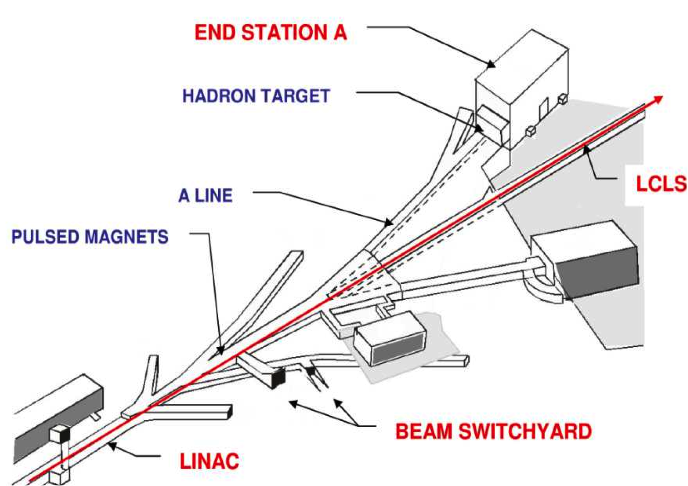}
\caption{SLAC accelerator complex at the end of the $3.2$~km linear accelerator.} \label{SLACschematic}
\end{figure*}

\begin{figure*}[htb]
\centering
\includegraphics[width=145mm]{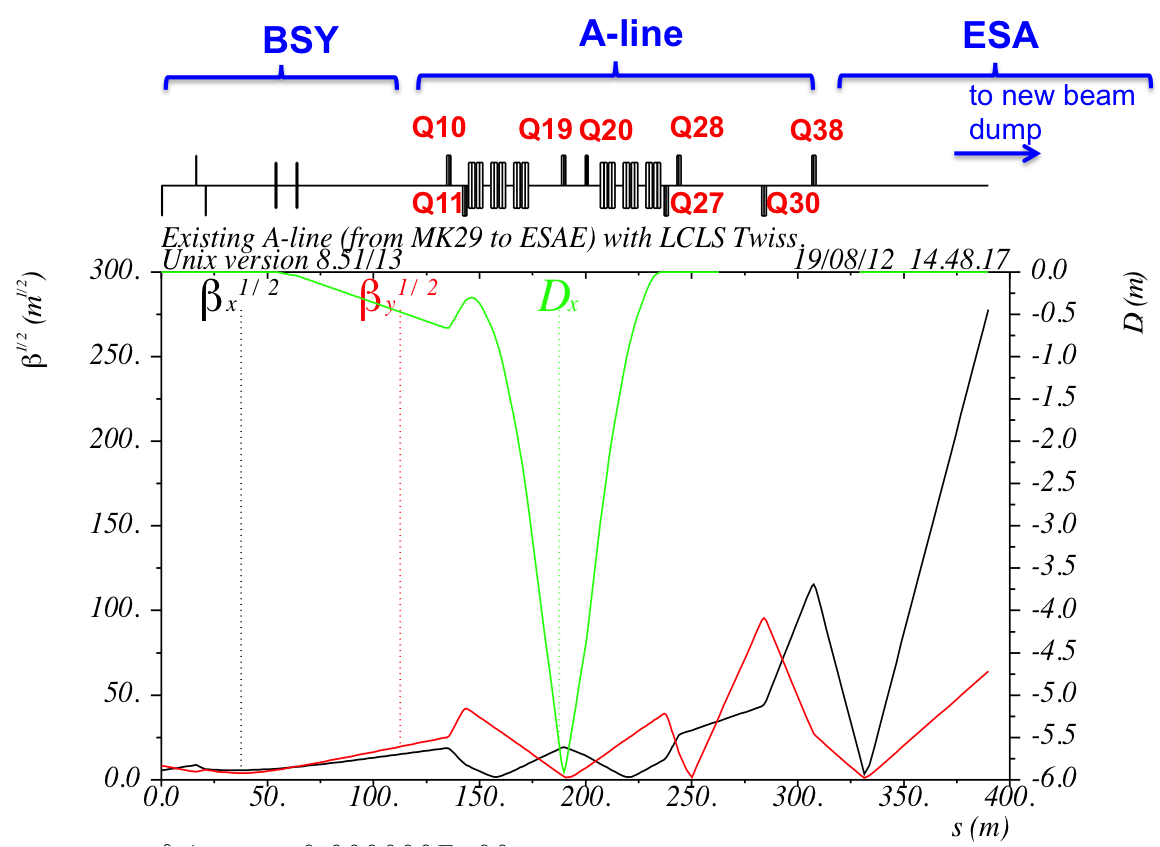}
\caption{Betatron and dispersion functions from the end of the Linac, through the BSY and
A-line to the east end of ESA.} \label{ESTBoptics}
\end{figure*}

End Station A Test Beam (ESTB) is a beam line at SLAC using a fraction of the bunches of the 15~GeV electron beam from the Linac Coherent Light Source (LCLS)\footnote{The LCLS at SLAC \cite{PEmma} is the world's first and most powerful X-ray  Free Electron Laser (FEL). Currently the LCLS operates in a range of beam energy $2.5$--$15$~GeV with a repetition rate of 120~Hz. Typically, its bunch charge ranges between 20 and 250 pC. With these beam features, the LCLS is able to provide X-rays in the wavelength range $1.5$--$15~\rm{\AA}$.}, restoring test beam capabilities in the End Station A (ESA) experimental hall. The infrastructure of these facilities is schematically shown in Fig.~\ref{SLACschematic}. 4 pulsed kicker magnets in the Beam Switch Yard (BSY) are used to kick the LCLS beam into ESA with a repetition rate of 5~Hz. There are possibilities of increasing the repetition rate when the beam is not  needed for LCLS operations, potentially doubling the available pulses to ESA. A complete description of the optics characteristics of ESTB is given in Ref.~\cite{ESTBSLAC}. Here we focus on optical issues relevant for the collimator wakefield experiment. The A-line optics has been matched to the LCLS beam characteristics in the BSY and an optical solution is shown in Fig.~\ref{ESTBoptics}.  

\begin{figure*}[htb]
\centering
\includegraphics[width=150mm]{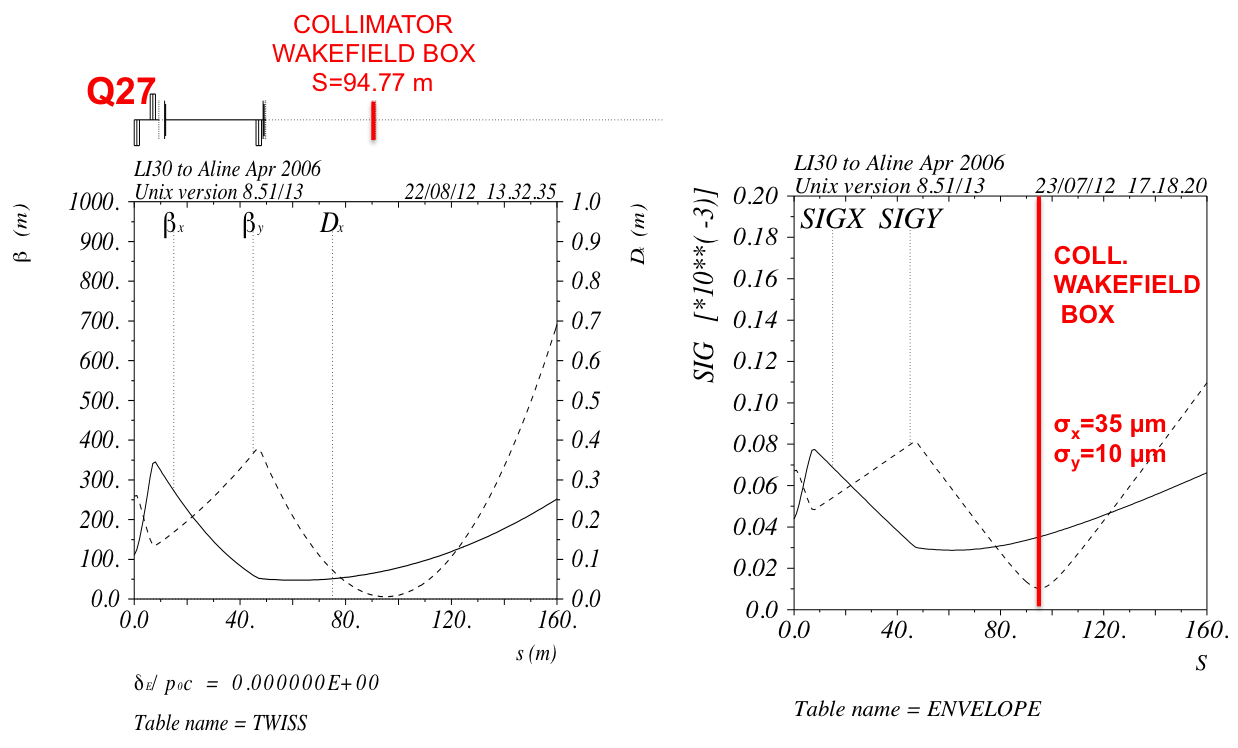}
\caption{Betatron functions (left) and the corresponding envelopes (right) through ESTB
from Q27. The position of the collimation wakefield box is indicated.} \label{ESTBzoom}
\end{figure*}

Four quadrupoles (Q27, Q28, Q30 and Q38) have been matched to obtain a vertical beam waist of $\sigma_y=10~\mu$m at the collimator box position. Figure~\ref{ESTBzoom} shows this optical solution from Q27, using the following initial twiss parameters at Q27 (measured with $10\%$ error on 26th April 2006): $\beta_{x0}=110.2$~m, $\alpha_{x0}=-4.9$, $\beta_{y0}=258.8$~m, $\alpha_{y0}=-6.0$.

\section{PRELIMINARY TRACKING SIMULATIONS}
\subsection{Wakefield Kick Factor}
\begin{figure*}[htb]
\centering
\includegraphics[width=130mm]{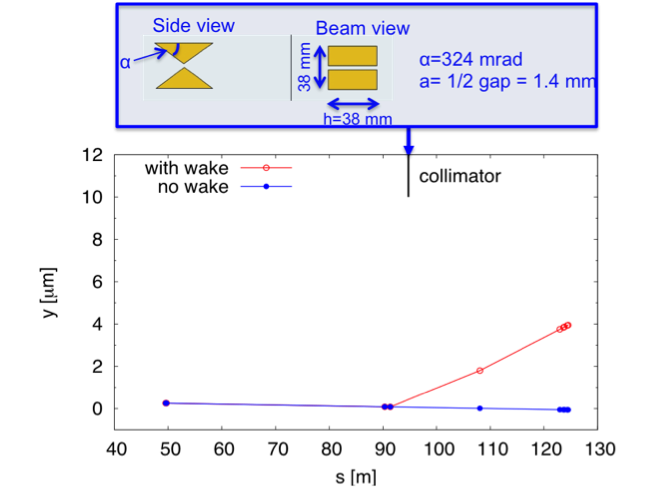}
\caption{Beam centroid trajectory through ESTB for a case of $-1.39$~mm vertical collimator displacement with respect to the nominal beam axis (case with wakefield effects), and for the case without collimator offset (reference orbit).} \label{ESTBtrajectories}
\end{figure*}

Preliminary beam tracking simulations, using the optics of Fig.~\ref{ESTBzoom}, have been performed in order to predict the level of kick angle due to collimator wakefield effects, and determine the position and resolution of BPMs. For these simulations we have considered the particular collimator configuration shown in Fig.~\ref{ESTBtrajectories}, where we compare the reference orbit and the deflected one due to wakefields, assuming a collimator offset w.r.t. beam axis of -1390 $\mu$m. In this example the trajectory has been reconstructed using 4 BPMs upstream of the collimator and 6 BPMs downstream of the collimator. In this particular case, a kick angle of about $0.15~\mu$rad has been obtained. These simulations have been performed using the tracking code PLACET, assuming a LCLS-like beam: $14.7$~GeV, $0.04\%$ energy spread, $\sigma_z=100~\mu$m bunch length, and $1.6\times 10^9$ electrons per bunch. Only the linear regime for the wakefield (up to quadrupolar component) has been considered.

\begin{figure*}[htb]
\centering
\includegraphics[width=100mm]{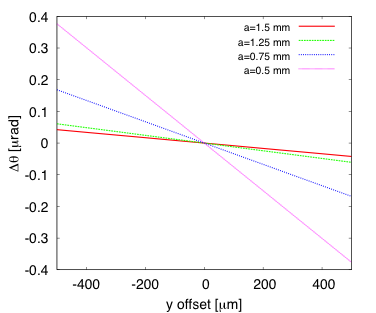}
\includegraphics[width=110mm]{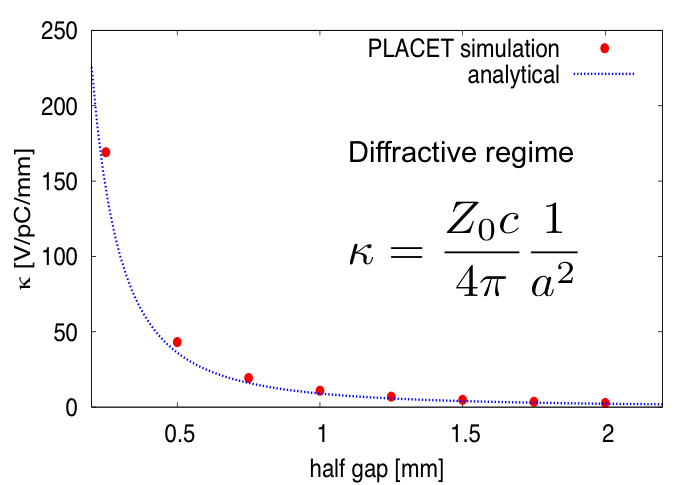}
\caption{Top: Deflection angle of the bunch centroid due to collimator wakefield effects for different collimator half gaps. Bottom: geometric kick factor as a function of the collimator half gap. The collimator geometry of Fig.~\ref{ESTBtrajectories} has been considered. } \label{kickangle}
\end{figure*}

Having the measurements of the trajectories one can obtain the deflection angle as a function of the beam-collimator separation (Fig.~\ref{kickangle}) and to calculate the transverse loss factor or kick factor from the linear fit of these curves. For the particular collimator geometry considered in this section, the geometric kick factor as a function of the collimator half gap is shown in Fig.~\ref{kickangle} (bottom), comparing the PLACET simulation results with the analytical calculation for the diffractive regime.

\subsection{Emittance Dilution}
\begin{figure*}[htb]
\centering
\includegraphics[width=100mm]{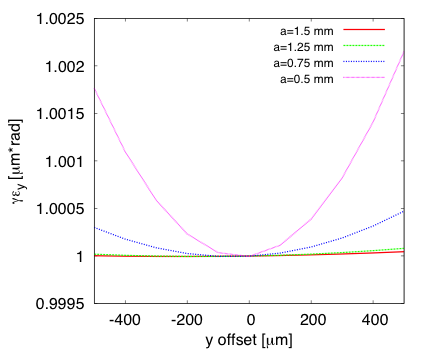}
\includegraphics[width=100mm]{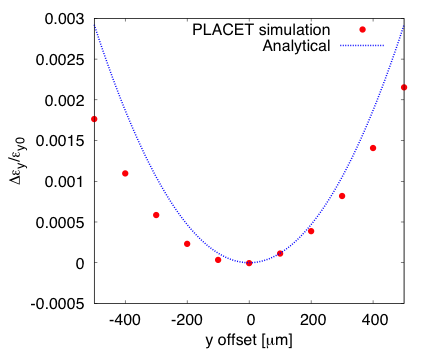}
\caption{Top: Vertical emittance as a function of the beam-collimator offset for different collimator half gaps. Bottom: Emittance growth for $a=0.5$~mm, comparing analytical results (from Eq.~(\ref{eq7})) with PLACET simulation results. The collimator geometry of Fig.~\ref{ESTBtrajectories} has been considered.} \label{emittdilution}
\end{figure*}

The emittance growth due to collimator wakefield effects can be calculated using the following expression: 

\begin{equation}\label{eq7}
\frac{\Delta \epsilon_y}{\epsilon_{y0}}=\sqrt{1+\frac{\beta_y}{\epsilon_{y0}}\langle y'^2_c \rangle}-1\,\,,
\end{equation}

\noindent where $\epsilon_{y0}$ is the design emittance, $\beta_y$ the betatron function at the collimator position, and $\langle y'^2_c \rangle$ the rms bunch centroid kick,

\begin{equation}\label{eq8}
\langle y'^2_c \rangle = \left( \frac{N_e r_e}{\gamma} \kappa^{\rm rms}_{\perp} y \right)^2 \,\,,
\end{equation}

\noindent where $\kappa^{\rm rms}_{\perp}$  is the spread of the transverse collimator wakefield kick. For a Gaussian bunch $\kappa^{\rm rms}_{\perp} =\kappa_{\perp}/\sqrt{3}$, with $\kappa_{\perp}$ the transverse wakefield kick factor in m$^{-2}$ units.  Figure~\ref{emittdilution} shows the emittance dilution due to collimator wakefield effects as a function of the beam-collimator offset considering the collimator configuration of Fig.~\ref{ESTBtrajectories}. If the collimator half gap $a=0.5$~mm, we expect an emittance dilution $\sim 0.1\%$ for beam offsets $\sim 100~\mu$m. Therefore, we expect a practically negligible effect on the emittance, and in this experiment we will basically focus on the study of the kick effect on the bunch centroid.

\subsection{BPM Resolution}
We require a measurement of the kick angle with $10\%$ resolution or better. Taking into account results from simulations above (Section 6.1), where kick angles of the order of $0.1~\mu$rad are expected to be generated by the wakefield effects, at least $0.01~\mu$rad resolution is required on the reconstructed kick angle. 

Simulations have been performed to determine the required BPM resolution for this experiment. In order to optimise the BPM positions in the ESTB beamline we have studied the following configurations: 

\begin{enumerate}
\item Let us consider the use of a BPM system such as it was used in the past for the ILC energy spectrometer experiment in ESA~\cite{ESAexperiment} (See Fig.~\ref{case1}). It consists of two BPM doublets upstream of the collimator wakefield box (BPMs 31, 32, and BPMs 1, 2) and two BPM triplets located downstream (BPMs 3, 4, 5, and BPMs 9, 10, 11). The distance between the BPMs in a same doublet or triplet is about $0.5$~m.

\begin{figure*}[htb]
\centering
\includegraphics[width=150mm]{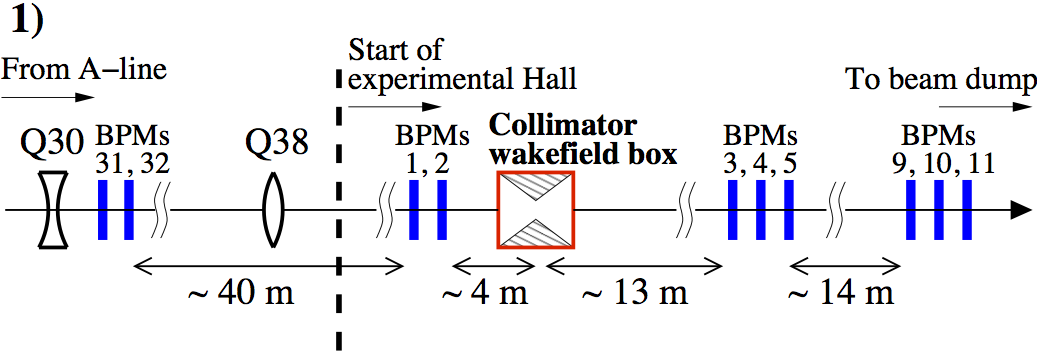}
\caption{Schematic showing the relative BPM positions in the ESTB beamline, assuming the same configuration as that of the ILC spectrometer experiment in the old ESA \cite{ESAexperiment}.} \label{case1}
\end{figure*}

\item Let us consider a case where we increase about 25~m the distance between the first and second BPM triplet with respect to the case 1.  See Fig.~\ref{case2}. For a precise measurement, the effect of the kick needs to be larger than the BPM resolution. Since a ballistic optical configuration is being used downstream of the collimator, the kick angle resolution $\sigma_{\rm kick}$ depend on the BPM resolution $\sigma_{\rm BPM}$ as follows: $\sigma_{\rm kick} \sim \sigma_{\rm BPM}/d$, where $d$ is the distance between the collimator and the BPM. Therefore, we could relax the BPM resolution constraint making the lever-arm large enough, i.e. increasing the distance $d$. In principle, this would allow to use BPMs with relatively poorer resolution. 

\begin{figure*}[htb]
\centering
\includegraphics[width=150mm]{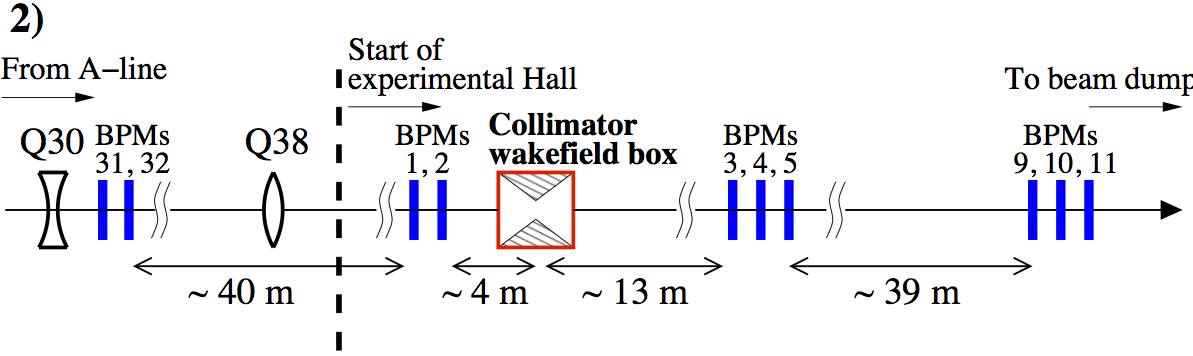}
\caption{Schematic showing the relative BPM positions in the ESTB beamline. In this case we have increased the distance between the two downstream BPM triplets with respect to the case 1. }\label{case2}
\end{figure*}

\item Considering a case similar to 1, but with the addition of another BPM triplet 25 m apart from BPM 11. See Fig.~\ref{case3}. In order to improve effective BPM resolution we can use a larger number of BPMs and/or averaging over a large number of pulses.

\begin{figure*}[htb]
\centering
\includegraphics[width=150mm]{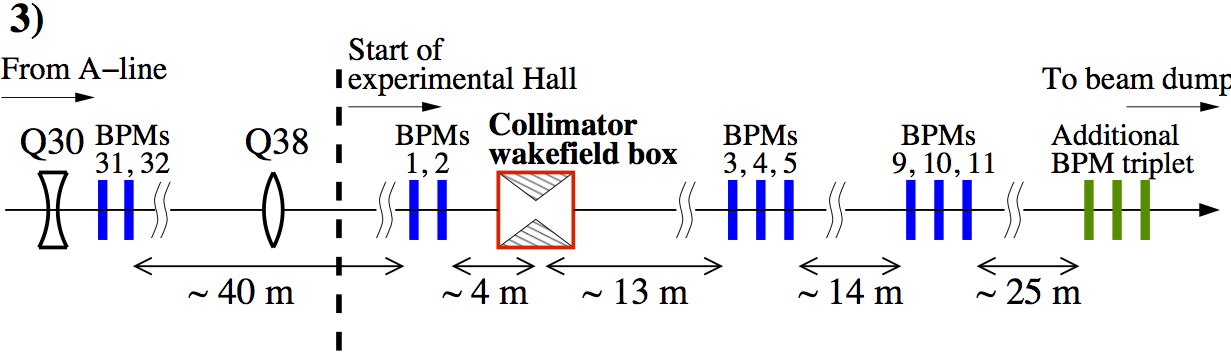}
\caption{Schematic showing the relative position BPM positions in the ESTB beamline. This case is similar to the case 1, but with one additional BPM triplet 25 m downstream of BPM 11.}\label{case3}
\end{figure*}

\end{enumerate}

\begin{figure*}[htb]
\centering
\includegraphics[width=120mm]{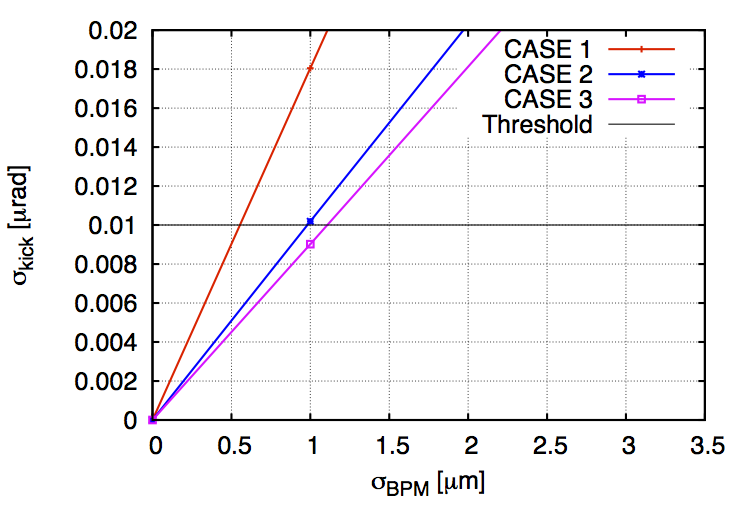}
\caption{Kick angle resolution as a function of the BPM resolution for the cases 1, 2 and 3 studied above.}\label{BPMresolutionplot}
\end{figure*}

Figure~\ref{BPMresolutionplot} shows the wakefield kick angle resolution as a function of the BPM resolution for the cases 1, 2 and 3. These results have been obtained from beam tracking simulations using the code PLACET. If a kick angle resolution $\sigma_{\rm kick} \approx 0.01~\mu$rad is required, then the necessary effective BPM resolution should be  $\sigma_{\rm BPM} \lesssim 0.6~\mu$m for case 1, $\sigma_{\rm BPM} \lesssim 1~\mu$m for case 2, and $\sigma_{\rm BPM} \lesssim 1.2~\mu$m for case 3. 

\section{PLANS}
\begin{itemize}
\item In a first phase of the experiment, we plan to measure collimator wakefields operating with a LCLS beam with bunch length in the range 100--300 $\mu$m. This study will be very useful to investigate collimator wakefields with ILC-like bunches ($\sigma_z=300~\mu$m), and for a precise benchmarking between experimental results, theory and simulations. 

For this experiment we plan to recycle instruments from the old ESA: wakefield box and BPMs. We are also working on the development of a new SPR monitor to reconstruct the bunch time profile and to measure the corresponding bunch length. 

\item In a second phase of the experiment, the plan is to push to smaller bunch length ($< 50~\mu$m) and to investigate collimator wakefields in regimes relevant for CLIC studies ($\sigma_z=44~\mu$m). 

The operation with short bunches is of interest for CLIC and FELs facilities. LCLS operates with ultra-short bunch length of $10~\mu$m and smaller. When the LCLS beam is diverted to the A-line by pulsed kickers to be used at ESTB (ESA), the bunch length increases to $100~\mu$m due to strong bends and large (optics) $R_{56}$. In order to reduce the bunch length in ESTB down to $20~\mu$m, a solution has been found by installing 4 additional quadrupoles in the A-line. The necessary funding to build and install these 4 quadrupoles will be determined by pending resources from interested parties. 

For the CLIC collimator studies we will require a SPR monitor in ESTB for precise bunch length measurements in the $\sim 100$~fs scale. Activities for the upgrade of the SPR monitor prototype, which is currently being tested in FACET, have already started to achieve reliable measurements of ultra-short bunch lengths.
\end{itemize}

\section{COLLABORATION}
\begin{itemize}
\item C. Hast, C. I. Clarke, SLAC, Stanford, USA.
\item J. Barranco, A. Latina, G. Rumolo, D. Schulte, R. Tom\'as, CERN, CH.
\item R. Appleby, R. M. Jones, Cockroft Institute, University of Manchester, UK.   
\item R. Bartolini, G. B. Christian, G. Doucas, I. Konoplev, C. Perry, A. Reichold, 
   A. Seryi, JAI, University of Oxford, UK.
\item  N. Delerue, LAL, Orsay, France.
\item    J. Smith, Lancaster University and Tech-X UK Ltd, UK.
\item   A. Faus-Golfe, N. Fuster-Mart\'inez, J. Resta-L\'opez, IFIC (CSIC-Valencia
   University), Spain.  
\end{itemize}


\end{document}